# Variable Earns Profit: Improved Adaptive Channel Estimation using Sparse VSS-NLMS Algorithms

Guan Gui[†], Linglong Dai[‡], Shinya Kumagai[†], and Fumiyuki Adachi[†]

[†]Department of Communications Engineering, Graduate School of Engineering, Tohoku University, Sendai 980-8579, Japan. E-mails: {gui, kumagai}@mobile.ecei.tohoku.ac.jp, adachi@ecei.tohoku.ac.jp
[‡]Tsinghua National Laboratory for Information Science and Technology, Department of Electronic Engineering, Tsinghua University, Beijing 10084, China. E-mail: daill@tsinghua.edu.cn

*Abstract*—Accurate channel estimation is essential for broadband wireless communications. As wireless channels often exhibit sparse structure, the adaptive sparse channel estimation algorithms based on normalized least mean square (NLMS) have been proposed, e.g., the zero-attracting NLMS (ZA-NLMS) algorithm and reweighted zero-attracting NLMS (RZA-NLMS). In these NLMS-based algorithms, the step size used to iteratively update the channel estimate is a critical parameter to control the estimation accuracy and the convergence speed (so the computational cost). However, invariable step-size (ISS) is usually used in conventional algorithms, which leads to provide performance loss or/and low convergence speed as well as high computational cost. To solve these problems, based on the observation that large step size is preferred for fast convergence while small step size is preferred for accurate estimation, we propose to replace the ISS by variable step size (VSS) in conventional NLMS-based algorithms to improve the adaptive sparse channel estimation in terms of bit error rate (BER) and mean square error (MSE) metrics. The proposed VSS-ZA-NLMS and VSS-RZA-NLMS algorithms adopt VSS, which can be adaptive to the estimation error in each iteration, i.e., large step size is used in the case of large estimation error to accelerate the convergence speed, while small step size is used when the estimation error is small to improve the steady-state estimation accuracy. Simulation results are provided to validate the effectiveness of the proposed scheme.

## I. INTRODUCTION

Broadband signal transmission is becoming one of the mainstream techniques in the next generation wireless communication systems [1][2]. The channel becomes severely frequency-selective and accurate channel state information (CSI) of such a channel is required for coherent detection (or demodulation). One of the effective approaches is the adaptive channel estimation (ACE) using normalized least mean square (NLMS) algorithm [3], which has low complexity and can be easily implemented at the receiver. On the other hand, many channel measurements have verified that wireless channels often exhibit a large delay spread but with a small nonzero taps support [4]–[6], and this channel sparsity has led to adaptive sparse channel estimation (ASCE) with improved accuracy. A typical example of sparse channel is shown in Fig. 1 with the length of finite impulse response (FIR) set to $N = 16$ and the number of dominant coefficients, $K = 3$. Unfortunately, ACE using NLMS algorithm always neglects the inherent sparse structure information. Hence, it may not be able to achieve the estimation performance comparable to sparse ASCE which exploits the channel sparsity. Inspired by least absolute shrinkage and selection operator (LASSO) algorithm [7], an $\ell_1$-norm sparse constraint function has been considered in the zero-attracting NLMS (ZA-NLMS) and reweighted ZA-NLMS (RZA-NLMS) algorithms to take advantage of the channel sparsity to improve the estimation performance.

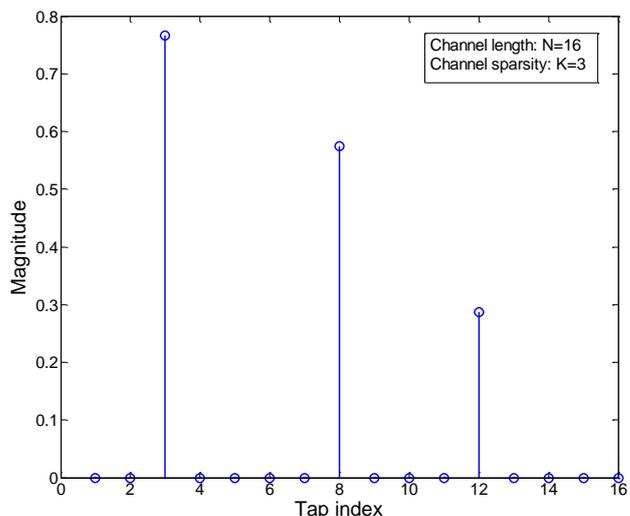

Fig. 1. A typical example of sparse multipath channel with channel length 16 and 3 nonzero taps.

It is well known that step size of the NLMS-based algorithms is a critical parameter to control the estimation performance, convergence speed and computational cost. In conventional NLMS-based algorithms including ZA-NLMS and RZA-NLMS, only invariable step size has been considered [3], which leads to provide performance loss or/and low convergence speed as well as high computational cost. Although variable step size NLMS (VSS-NLMS) was proposed for ACE to improve the estimation accuracy [8], channel sparsity has not considered in the VSS-NLMS algorithm.

In this paper, by jointly taking advantage of channel sparsity and VSS-NLMS, we propose two improved ASCE methods named as variable step size zero-attracting NLMS (VSS-ZA-NLMS) and VSS reweighted ZA-NLMS (VSS-RZA-NLMS) algorithms. Based on the observation that large step size is preferred for fast convergence while small step size is preferred for accurate estimation, the proposed VSS-ZA-NLMS and VSS-RZA-NLMS algorithms replace the ISS by VSS in conventional NLMS-based algorithms to improve the adaptive sparse channel estimation in terms of bit error rate (BER) and mean square error (MSE) metrics. The VSS is adaptive to the estimation error in each iteration, i.e., large step size is used in the case of large estimation error to accelerate the convergence speed, while small step size is used when the estimation error is small to improve the steady-state estimation accuracy.

The remainder of the rest paper is organized as follows. A system model is described at first and then the drawback of sparse channel estimation using sparse ISS-NLMS algorithms is introduced in Section II. In section III, sparse VSS-NLMS algorithms are proposed for improving accuracy of the channel estimators. Computer simulation results are presented in Section IV in order to compare the performances of the proposed algorithms. Finally, we conclude the paper in Section V.

*Notation*: Throughout the paper, matrices and vectors are represented by boldface upper case letters and boldface lower case letters, respectively; the superscripts $(\cdot)^T$, $(\cdot)^H$, $\text{Tr}(\cdot)$ and $(\cdot)^{-1}$ denote the transpose, the Hermitian transpose, the trace and the inverse operators, respectively; $E\{\cdot\}$ denotes the expectation operator; $\|\mathbf{h}\|_0$ is the $\ell_0$-norm operator that counts the number of nonzero taps in $\mathbf{h}$ and $\|\mathbf{h}\|_p$ stands for the $\ell_p$-norm operator which is computed by $\|\mathbf{h}\|_p = (\sum_i |h_i|^p)^{1/p}$, where $p \in \{1,2\}$ is considered in this paper; $\text{sgn}(\cdot)$ is a component-wise function which is defined by $\text{sgn}(h) = 1$ for $h > 0$, $\text{sgn}(h) = 0$ for $h = 0$, and $\text{sgn}(h) = -1$ for $h < 0$.

## II. DRAWBACKS OF ADAPTIVE CHANNEL ESTIMATION USING SPARSE ISS-NLMS ALGORITHMS

Consider a baseband-equivalent frequency-selective fading wireless communication system where the sparse channel vector $\mathbf{h} = [h_0, h_1, \ldots, h_{N-1}]^T$ of length $N$ has only $K$ nonzero channel taps. Assume that an input training signal $x(t)$ is used to probe the unknown sparse channel. At the receiver side, the corresponding observed signal $y(t)$ is given by

$$y(t) = \mathbf{h}^T \mathbf{x}(t) + z(t), \quad (1)$$

where $\mathbf{x}(t) = [x(t), x(t-1), \ldots, x(t-N+1)]^T$ denotes the vector of input signal $x(t)$; $z(t)$ is an additive white Gaussian noise (AWGN), which is assumed to be independent with $x(t)$; The objective of ASCE is to adaptively estimate the unknown sparse channel vector $\mathbf{h}$ using the training signal vector $\mathbf{x}(t)$ and the observed signal $y(t)$.

### A. ISS-ZA-NLMS algorithm

By defining the square estimation error at the $n$-th update by $e^2(n)$, ISS-ZA-NLMS [9] was proposed as

$$\hbar(n+1) = \hbar(n) + \mu \frac{e(n)\mathbf{x}(t)}{\mathbf{x}^T(t)\mathbf{x}(t)} - \rho_{ZA} \text{sgn}(\hbar(n)). \quad (2)$$

where $\hbar(n)$ is the $n$-th iterative adaptive channel estimator; $\mu \in (0, 1/\lambda_{\max})$ is the ISS; $\lambda_{\max}$ is the maximum eigenvalue of $\mathbf{R} = E[\mathbf{x}(t)\mathbf{x}^T(t)]$; and $\rho_{ZA} = \mu \lambda_{ZA}$ is a parameter which depends on the ISS and sparse regularization parameter. The third term $\rho_{ZA} \text{sgn}(\hbar(n))$ is used to attract small channel coefficients as zero in a high probability. In other words, most of small channel coefficients can be replaced directly by zeros, which is helpful to speed up the convergence of this algorithm and also to mitigate the extra noise interference on zero positions.

### B. ISS-RZA-NLMS algorithm

The sparse constraint of ISS-ZA-NLMS always gives the identical penalty to all the taps which are forced to be zero with the same probability. Motivated by the reweighted $\ell_1$-norm minimization recovery algorithm [10], an improved algorithm ISS-RZA-NLMS [9] was proposed as

$$\hbar(n+1) = \hbar(n) + \mu e(n)\mathbf{x}(t) - \rho_{\text{RZA}} \frac{\text{sgn}(\hbar(n))}{1 + \varepsilon_{RZA}|\hbar(n)|}. \quad (3)$$

where $\rho_{\text{RZA}} = \mu \lambda_{RZA} \varepsilon_{RZA}$; $\lambda_{RZA}$ is the regularization parameter and $\varepsilon_{RZA}$ is the reweighted factor which is set as $\varepsilon_{RZA} = 20$ as suggested in [11]. Please note that the third term in Eq. (3) attracts the channel coefficients $\tilde{h}_i(n)$, $n = 0,1,\ldots,N-1$ whose magnitudes are comparable to $1/\varepsilon_{RZA}$ to zeros.

### C. Drawbacks of the two sparse ISS-NLMS algorithms

Comparing the standard ISS-NLMS algorithm [3], sparse ISS-NLMS algorithms have a common ability of exploiting channel sparsity. Without the loss of generality, we derive the steady-state mean square error (MSE) performance of the ISS-ZA-NLMS [9] as for the typical example to illustrate the drawbacks of the sparse ISS-NLMS algorithms. Under the independence assumption, in [12], the steady-state MSE of ISS-NLMS estimator $\hbar_s(n)$ was derived as

$$\xi_t(\infty) = \lim_{n \to \infty} E\left\{\left[(\hbar(n) - \mathbf{h})^T \mathbf{x}(t)\right]^2\right\}$$

$$= \frac{\text{Tr}[\mathbf{R}(\mathbf{I} - \mu\mathbf{R})^{-1}]\sigma_n^2}{2 - \text{Tr}[\mathbf{R}(\mathbf{I} - \mu\mathbf{R})^{-1}]} \geq \frac{\lambda_{\max} \sigma_n^2}{2 - 3\mu\lambda_{\max}}, \quad (4)$$

where $\sigma_n^2$ is noise power of the $z(t)$. Similarly, steady-state MSE of the ISS-ZA-NLMS estimator $\hbar(n)$ was also derived as

$$\xi_s(\infty) = \lim_{n \to \infty} E\left\{\left[(\hbar(n) - \mathbf{h})^T \mathbf{x}(t)\right]^2\right\}$$

$$= \frac{\text{Tr}[\mathbf{R}(\mathbf{I} - \mu\mathbf{R})^{-1}]\sigma_n^2}{2 - \text{Tr}[\mathbf{R}(\mathbf{I} - \mu\mathbf{R})^{-1}]} + \frac{\gamma_1 \rho_{\text{ZA}} (\rho_{\text{ZA}} - 2\gamma_2/\gamma_1)}{(2 - \text{Tr}[\mathbf{R}(\mathbf{I} - \mu\mathbf{R})^{-1}])\mu}$$

$$< \frac{\text{Tr}[\mathbf{R}(\mathbf{I} - \mu\mathbf{R})^{-1}]\sigma_n^2}{2 - \text{Tr}[\mathbf{R}(\mathbf{I} - \mu\mathbf{R})^{-1}]}$$

$$\leq \frac{\lambda_{\max} \sigma_n^2}{2 - 3\mu\lambda_{\max}} \leq \xi_t(\infty), \quad (5)$$

where $\gamma_1 = E\left[\text{sgn}(\hbar^T(n))(\mathbf{I} - \mu\mathbf{R})^{-1}\text{sgn}(\hbar(n))\right] > 0$ and $\gamma_2 = E\|\hbar(\infty)\|_1 - \|\mathbf{h}\|_1$. To exploit the channel sparsity, $\rho_{\text{ZA}}$ should be selected in the range $(0, 2\gamma_2/\gamma_1)$ so that $(\rho_{\text{ZA}} -$

$2\gamma_2/\gamma_1) \leq 0$. Hence, $\xi_s(\infty)$ in Eq. (5) is lower than $\xi_t(\infty)$ in Eq. (4). According to Eq. (5), the lower bound of $\xi_s(\infty)$ depends on the three factors: $\{\lambda_{max}, \sigma_n^2, \mu\}$. However, $\lambda_{max}$ and $\sigma_n^2$ are decided by the input signal $\mathbf{x}(t)$ and additive noise $z(t)$, respectively. Only selecting the smaller step-size can further achieve better MSE performance. However, if small step-size $\mu$ is adopted, it will incur slow convergence speed (i.e., high computation complexity) on overall adaptive channel estimation. Hence, it is expected that large step-size is used in the case of large MSE to accelerate the convergence speed, while small step-size is used in the case of smaller MSE to improve the steady-state MSE performance. As $\mu \to 0$, the lower bound of steady-state MSE of sparse ISS-NLMS algorithm is derived as

$$\lim_{\mu \to 0} \xi_s(\infty) \leq \lim_{\mu \to 0} \frac{\lambda_{max}\sigma_n^2}{2 - 3\mu\lambda_{max}} = \frac{\lambda_{max}\sigma_n^2}{2}. \quad (6)$$

To simultaneously achieve higher convergence speed and lower steady-state MSE performance, we propose sparse VSS-NLMS algorithms in the next section.

## III. Improved Adaptive Channel Estimation using Sparse VSS-NLMS Algorithms

Recall that the sparse ISS-NLMS algorithms in Eqs. (2) and (3) does not utilize VSS. Inspired by the VSS-NLMS algorithm which has been proposed in [8], to further improve the estimation performance, VSS is introduced to adapt to the changes of the estimation error. For the given observed signal $y(t)$, based on the previous research on ISS-ZA-NLMS algorithm in Eq. (2), VSS-ZA-NLMS algorithm performs as follows

$$\tilde{\mathbf{h}}(n+1) = \tilde{\mathbf{h}}(n) + \mu(n+1)\frac{e(n)\mathbf{x}(t)}{\mathbf{x}^T(t)\mathbf{x}(t)} - \rho_{ZA}\operatorname{sgn}(\tilde{\mathbf{h}}(n)), \quad (7)$$

where $\mu(n+1)$ is the VSS, which is updated as

$$\mu(n+1) = \mu_{max} \cdot \frac{\mathbf{p}^T(n+1)\mathbf{p}(n+1)}{\mathbf{p}^T(n+1)\mathbf{p}(n+1) + C}, \quad (8)$$

where $C$ is a positive threshold parameter which is related to $\sigma_n^2 \operatorname{Tr}\{[\mathbf{x}(t)\mathbf{x}^T(t)]^{-1}\}$ and can be written as $C \sim \mathcal{O}(1/\mathrm{SNR})$, where SNR is the received signal-to-noise ratio (SNR). According to Eq. (7), the range of VSS is given by $\mu(n+1) \in (0, \mu_{max})$, where $\mu_{max}$ is the maximal step-size. To ensure the stability of the adaptive algorithm (7), the maximal step-size is less than 2 [3]. It is worth mentioning that $\mathbf{p}(n)$ in Eq. (8) is defined by

$$\mathbf{p}(n+1) = \beta\mathbf{p}(n) + (1-\beta)\frac{\mathbf{x}(t)e(n)}{\mathbf{x}^T(t)\mathbf{x}(t)}, \quad (9)$$

where $\beta \in [0,1]$ is the smoothing factor for controlling the VSS and estimation error.

Similarly, based on the conventional ISS-RZA-NLMS algorithm (3), a update version called as VSS-RZA-NLMS algorithm is proposed as follows,

$$\tilde{\mathbf{h}}(n+1) = \tilde{\mathbf{h}}(n) + \mu(n+1)\frac{e(n)\mathbf{x}(t)}{\mathbf{x}^T(t)\mathbf{x}(t)} - \rho_{RZA}\frac{\operatorname{sgn}(\tilde{\mathbf{h}}(n))}{1 + \varepsilon_{RZA}|\tilde{\mathbf{h}}(n)|}. \quad (11)$$

Based on the sparse VSS-NLMS algorithms in Eqs. (7) and (11), two improved adaptive sparse channel estimation methods are summarized as in Tab. I.

TAB. I. SPARSE VSS-NLMS ALGORITHMS FOR ESTIMATING CHANNELS.

| | |
|---|---|
| Input | training signal vector: $\mathbf{x}(t)$ |
| | observed signal: $y(t)$ |
| | initial step-size: $\mu_{max}$ |
| | sparse regularization parameter: $\lambda_{(R)ZA}$ |
| | positive threshold parameter: $C$ |
| | smoothing factor: $\beta$ |
| | (*) reweighted factor of VSS-RZA-NLMS: $\varepsilon_{RZA}$ |
| Output | channel estimator $\tilde{\mathbf{h}}(n+1)$ |
| Step 1 Initialize | update times: $n = 1$ |
| | initial adaptive channel estimator: $\tilde{\mathbf{h}}(0) = 0$ |
| | initial projection vector: $\mathbf{p}(0) = 0$ |
| Step 2 Compute error | $e(n) = y(t) - \tilde{\mathbf{h}}^T(n-1)\mathbf{x}(n)$ |
| Step 3 Set adaptive VSS | Compute the adaptive VSS by according the Eq. (8) |
| Step 4 Adaptive updating | Adaptive channel estimation using algorithms: VSS-ZA-NLMS in Eq. (7) or VSS-RZA-NLMS in Eq. (11) |
| Step 5: Stop criterion | If $\|\tilde{\mathbf{h}}(n+1) - \tilde{\mathbf{h}}(n)\|_2^2 \leq 10^{-5}$ or $n \geq 5000$ is satisfied, then terminate the algorithm and output $\tilde{\mathbf{h}}(n+1)$; otherwise, $n = n+1$, run the algorithm from **step 2** to **step 5**. |

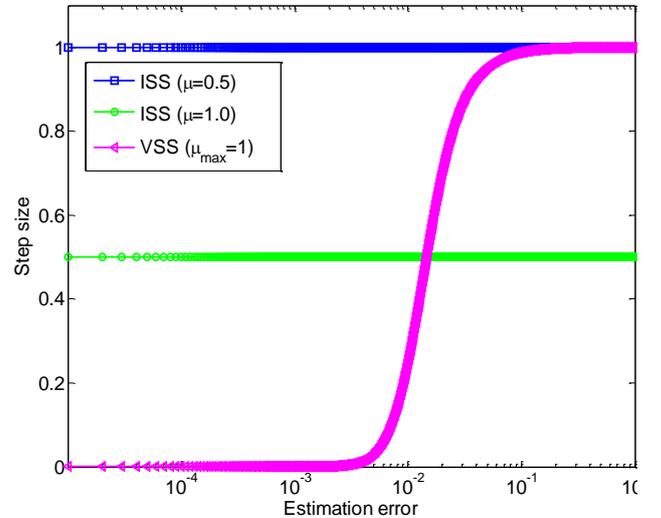

Fig. 2. Step size vs. estimation error.

For a better understanding of the difference between ISS and VSS, step size $\mu$ for sparse ISS-NLMS algorithms is invariable but the step size $\mu(n+1)$ for sparse VSS-NLMS algorithms is variable as depicted in Fig. 2, where the maximal step size $\mu_{\max}$ and step size $\mu$ are set as $\mu_{max} = 1$ and $\mu \in \{0.5,1\}$, respectively. From Fig. 2, one can easily find that for VSS, $\mu(n)$ decreases as the estimation performance increases and vice versa, while the step size $\mu$ in conventional ISS-NLMS algorithms are kept invariant.

## IV. COMPUTER SIMULATIONS

To validate the effectiveness of the proposed method, two metrics, i.e., MSE and BER, are adopted. Channel estimators are evaluated by average MSE which is defined by

$$\text{Average MSE}\{\hbar(n)\} = \text{E}\left\{\|\mathbf{h} - \hbar(n)\|_2^2\right\}, \quad (12)$$

System performance is evaluated in terms of BER which adopts different data modulation schemes. The results are averaged over 1000 independent Monte-Carlo runs. Each dominant channel tap follows random Gaussian distribution as $\mathcal{CN}(0, \sigma_h^2)$ which is subject to $\text{E}\{\|\mathbf{h}\|_2^2 = 1\}$ and their positions are randomly decided within the $\mathbf{h}$. The received SNR is defined as $P_0/\sigma_n^2$, where $P_0$ is the received power of the pseudo-random noise (PN)-sequence for training signal. In addition, to achieve better steady-state estimation performance, reweighted factor of sparse RZA-NLMS algorithms (using ISS and VSS) is set as $\varepsilon_{RZA} = 20$ [11][13]. Parameters for computer simulation are given in Tab. II.

TAB. II. SIMULATION PARAMETERS.

| Parameters | Values |
|---|---|
| Channel length | $N = 60$ |
| No. of nonzero coefficients | $K = 3$ and 6 |
| Distribution of nonzero coefficient | Random Gaussian $\mathcal{CN}(0,1)$ |
| Threshold parameter for VSS-NLMS | $C = 10^{-4}$ for 5dB<br>$C = 10^{-5}$ for {10dB,20dB} |
| Received SNR for channel estimation | {5dB,10dB,20dB} |
| Received SNR $E_s/N_0$ for symbol | 12dB~30dB |
| Step-size of gradient descend | $\mu = 0.2$ and $\mu_{\max} = 2$ |
| Regularization parameters for sparse penalties | $\rho_{ZA} = 0.0002\sigma_n^2$<br>$\rho_{RZA} = 0.002\sigma_n^2$ |
| Modulation schemes | 8PSK,16PSK,16QAM, 64QAM |

In the first example, average MSE performance of the proposed method are evaluated for $K = 3$ and 6 in Figs. 3-4 under three SNR regimes, i.e. 5dB, 10dB and 20dB. To confirm the effectiveness of the proposed method, they are compared with three previous methods, i.e., ISS-NLMS [3], VSS-NLMS [8] and sparse ISS-NLMS algorithms [14] [9]. In the case of different SNR regimes, sparse VSS-NLMS algorithms always achieve better performance with respect to average MSE while faster convergence speed with respect to iteration times than sparse ISS-NLMS ones. Because sparse VSS-NLMS algorithms utilize the step-size which is time-variant relates to estimation error. In other word, sparse VSS-NLMS algorithm adopts small step-size adaptively to achieve better performance while utilizing a large step-size to improve convergence rate. In addition, sparse VSS-NLMS algorithms also take advantage of the channel sparsity, they obtain a better estimation performance than sparse ISS-NLMS, especially in the extreme sparse channel case, e.g., $K = 3$, as shown in Figs. 3(a) and 4(a).

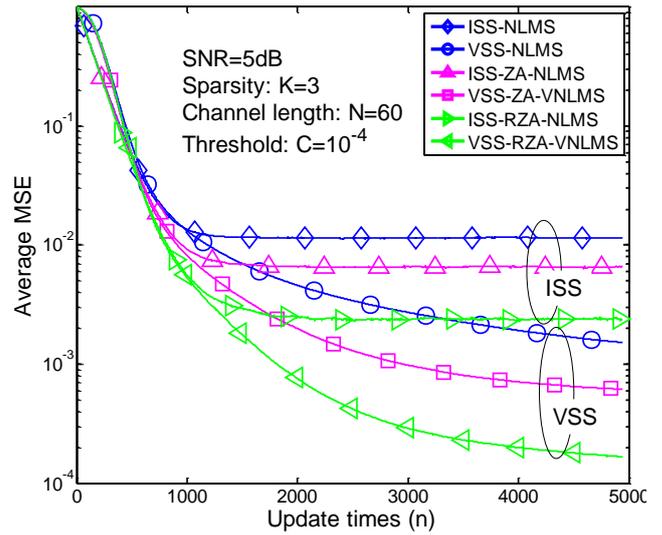

(a) SNR=5dB and K=3

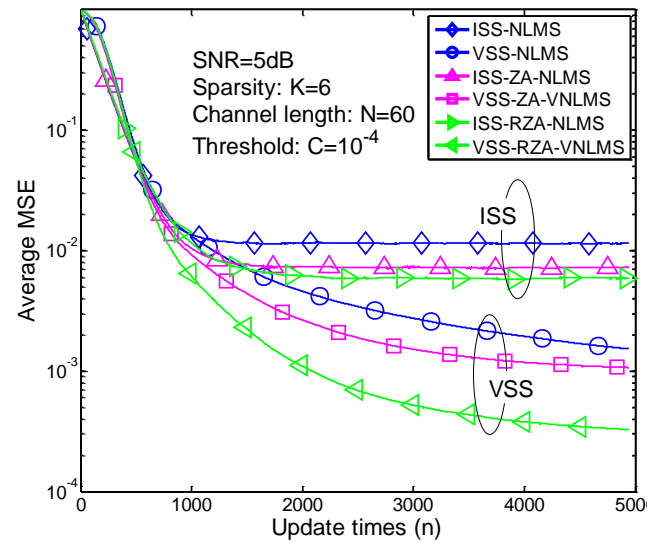

(b) SNR=5dB and K=5

Fig. 3. Average MSE versus algorithm update times.

In the second example, the average BER performance using proposed channel estimators is also evaluated. Assume $K = 3$ (the number of nonzero taps of the channel), the steady-state channel estimator (5dB) is adopted for evaluating the system performance. Indeed, the evaluation of exact BER can be quite cumbersome because it depends on the bit-to-symbol mapping used [15]. To avoid the high computation, here, a simple BER evaluation method via invertible exponential-type approximations is adopted [15]. For the multilevel phase shift keying (PSK) modulation and multilevel quadrature amplitude

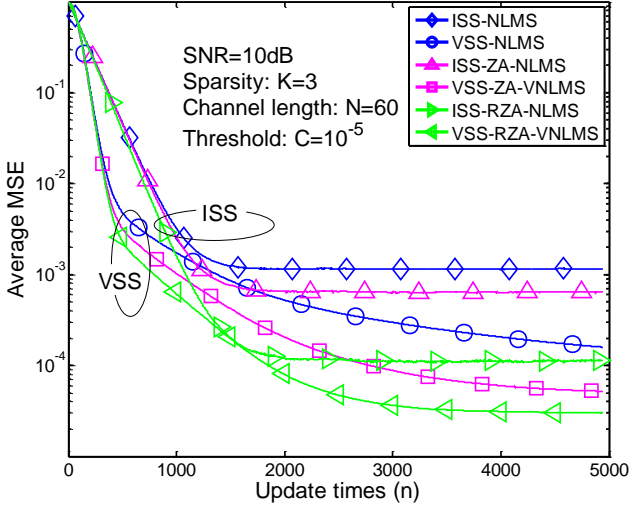

(a) SNR=10dB and K=3.

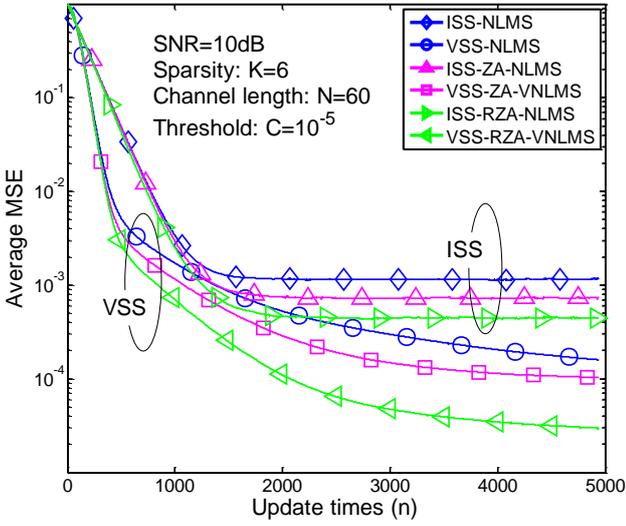

(b) SNR=10dB and K=6.

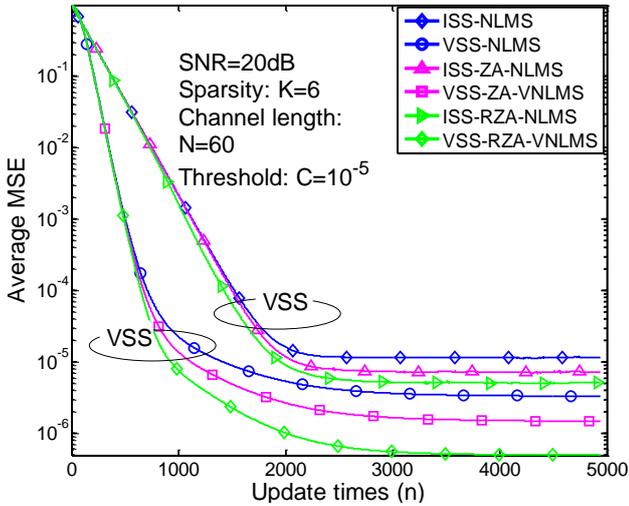

(c) SNR=20dB and K=6.

Fig. 4. Average MSE versus algorithm update times.

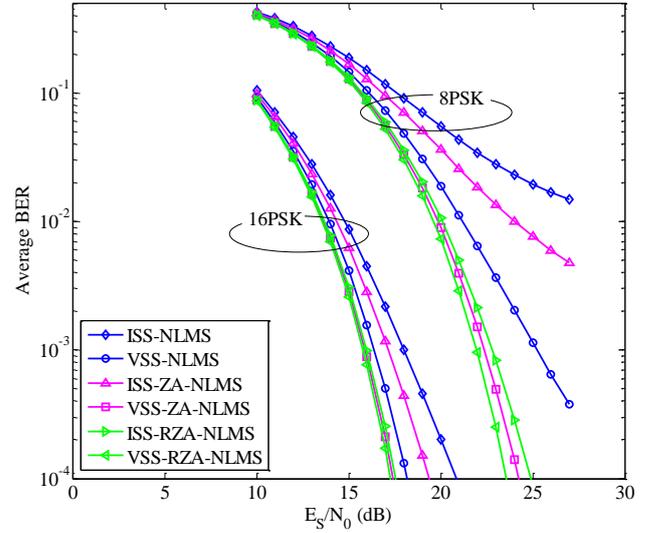

Fig. 5. Average BER performance versus SNR with respect to PSK.

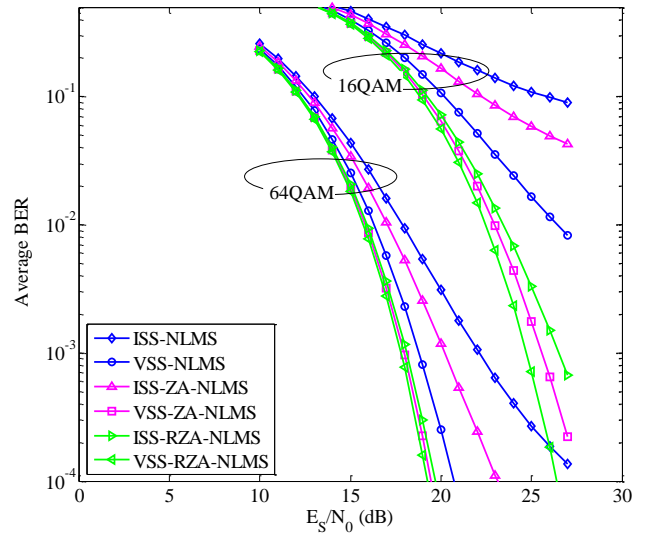

Fig. 6. Average BER performance versus SNR with respect to QAM.

modulation (QAM) schemes, their average BERs can be computed by

$$a_1 e^{-b\gamma_s \sin^2(\pi/M)} + a_2 e^{-2b\gamma_s \sin^2(\pi/M)}, \quad M \geq 4 \quad (13)$$

and

$$2ka_1 e^{-\frac{1.5b\gamma_s}{M-1}} + (2ka_2 - k^2 a_1^2) e^{-\frac{3b\gamma_s}{M-1}} - k^2 a_2^2 e^{-\frac{6b\gamma_s}{M-1}}$$
$$-2k^2 a_1 a_2 e^{-\frac{4.5b\gamma_s}{M-1}}, \text{for } k = (\sqrt{M}-1)/\sqrt{M}, \quad (14)$$

respectively, where $a_1 = 0.3017$, $a_2 = 0.438$, and $b = 1.0510$ are the optimal curve-fitting coefficients; $M$ is multilevel of modulation and $\gamma_s$ is defined as

$$\gamma_s = \frac{10^{\frac{SNR}{10}}(1-MSE)}{10^{\frac{SNR}{10}} MSE + 1}. \quad (15).$$

Received SNR is defined by $E_S/N_0$, where $E_S$ is the received power of symbol and $N_0$ is the noise power. In Fig. 5, the average BER performances of multilevel PSK modulation schemes, i.e., 8PSK and 16PSK, are plotted as a function of $E_S/N_0$. One can find that both VSS-ZA-NLMS and VSS-RZA-NLMS can achieve better estimation accuracy than ISS-ZA-NLMS and ISS-RZA-NLMS algorithms. In addition, the steady-state channel estimator of VSS-RZA-NLMS is better than the one of VSS-ZA-NLMS due to the fact that the former algorithm takes more sparse information than latter one. In Fig. 6, multilevel QAM schemes, i.e., 16QAM and 64QAM, are considered for data modulation. It is observed that the proposed method can achieve a better estimation than previous methods. It is expected that the BER performance could also be improved when considering channel coding techniques

## V. Conclusion

The drawback of sparse ISS-NLMS based algorithms is that they cannot balance the convergence speed and steady-state performance for adaptive sparse channel estimation. Unlike the traditional algorithms, in this paper, we proposed two sparse VSS-NLMS algorithms, i.e., VSS-ZA-NLMS and VSS-RZA-NLMS to improve adaptive estimation accuracy. The proposed algorithms utilize VSS which can change adaptively the estimation error, i.e, the step-size becomes smaller as the estimation accuracy improves and vice versa. Simulation results were provided to validate the effectiveness of the proposed algorithms in terms of MSE and BER.

## Acknowledgments

This work was supported by grant-in-aid for the Japan Society for the Promotion of Science (JSPS) fellows grant number 24·02366 and National Key Basic Research Program of China (Grant No. 2013CB329203), National Natural Science Foundation of China (Grant Nos. 61271266 and 61201185).